\documentclass[a4paper,10pt,twoside]{cpc-hepnp}
\usepackage{CJK,upgreek,fancyhdr}
\usepackage{multicol}
\usepackage{graphicx}
\usepackage{booktabs}
\usepackage{amssymb,bm,mathrsfs,bbm,amscd}
\usepackage[tbtags]{amsmath}
\usepackage{lastpage}

\usepackage{color}
\usepackage[normalem]{ulem}%Include underunderline using uuline{...}, underwave using uwave{...},line-through using \sout{...}  Note that ulem changes the behavior of emph. In order to avoid it, add \normalem to the preamble after the package insertion.

\usepackage{slashed}
\usepackage{tabularx}

\usepackage{natbib}
\usepackage{amssymb}
\usepackage{amsmath}
\usepackage{amsfonts}
\usepackage{epsfig}
\usepackage{graphicx}
\usepackage{tabularx}
\usepackage{latexsym}
\usepackage{subfigure}
\usepackage{verbatim}
\usepackage{color}
\usepackage{braket}
\usepackage{multirow}
\usepackage{dcolumn}% Align table columns on decimal point
\usepackage{placeins}
\usepackage{epstopdf}
\usepackage{wrapfig}
\usepackage{hyperref}
\usepackage{booktabs}
\usepackage{simplewick}
\usepackage[normalem]{ulem}
\usepackage[toc,page]{appendix}
\renewcommand\sout{\bgroup \color{red} \ULdepth=-.5ex \ULset}

\def\eq{\begin{eqnarray}}
\def\en{\end{eqnarray}}

\begin{document}
\begin{CJK*}{GBK}{song}

\fancyhead[c]{\small Chinese Physics C~~~Vol. xx, No. x (20xx)
xxxxxx} \fancyfoot[C]{\small xxxxxx-\thepage}

\footnotetext[0]{Received xxx}

\title{Constraining the anisotropy of the Universe via Pantheon supernovae sample\thanks{Supported by National Natural Science
Foundation of China (11675182, 11690022).}}

\author{%
      Zhe Chang$^{1,2}$%
\quad Dong Zhao$^{1,2}$%
\quad Yong Zhou$^{1,2;1)}$\email{zhouyong@ihep.ac.cn}
}
\maketitle

\address{%
$^{1}$Institute of High Energy Physics, Chinese Academy of Sciences, Beijing 100049, China\\
$^{2}$School of Physical Sciences, University of Chinese Academy of Sciences, Beijing 100049, China
}

\begin{abstract}
We test the possible dipole anisotropy of a Finslerian cosmological model and other three dipole-modulated cosmological models, i.e., the dipole-modulated $\rm{\Lambda}$CDM, $w$CDM and Chevallier--Polarski--Linder (CPL) model by using the recently released Pantheon sample of SNe Ia. The Markov chain Monte Carlo (MCMC) method is used to explore the whole parameter space. We find that the dipole anisotropy is very weak in all cosmological models used. Although the dipole amplitudes of four cosmological models are consistent with zero within $1\sigma$ uncertainty, the dipole directions are close to the axial direction to the plane of SDSS subsample among Pantheon. It may imply that the weak dipole anisotropy in the Pantheon sample originates from the inhomogeneous distribution of the SDSS subsample. More homogeneous distribution of SNe Ia is necessary to constrain the cosmic anisotropy.
\end{abstract}


\begin{pacs}
97.60.Bw, %
98.65.Dx, %
98.80.k
%95.30.Sf	Relativity and gravitation (see also section 04 General relativity and gravitation; 98.80.Jk Mathematical and relativistic aspects of cosmology)
%98.52.Nr	Spiral galaxies
%98.65.Dx	Superclusters; large-scale structure of the Universe (including voids, pancakes, great wall, etc.)
%98.62.Ck	Masses and mass distribution
%98.62.Dm	Kinematics, dynamics, and rotation
%98.80.-k	Cosmology
%1---3 PACS codes (Physics and Astronomy Classification Scheme, http://www.aip.org/pacs/pacs.html/)
\end{pacs}

\footnotetext[0]{\hspace*{-3mm}\raisebox{0.3ex}{$\scriptstyle\copyright$}2019
Chinese Physical Society and the Institute of High Energy Physics
of the Chinese Academy of Sciences and the Institute
of Modern Physics of the Chinese Academy of Sciences and IOP Publishing Ltd}%

\begin{multicols}{2}

\section{Introduction}\label{sec:introduction}
The cosmological principle assumes that our Universe is homogeneous and isotropic on large scale, which is one of the foundations of modern cosmology \cite{Weinberg:2008zzc}. During the past few decades, the cosmological principle had been tested many times and found to be well consistent with most cosmological observations, for instance, the halo power spectrum \cite{Reid:2009xm}, the statistics of galaxies \cite{TrujilloGomez:2010yh}, the cosmic microwave background (CMB) radiation from \textit{Wilkinson Microwave Anisotropy Probe} (WMAP) \cite{Bennett:2012zja,Hinshaw:2012aka} and \textit{Planck} satellites \cite{Ade:2013sjv,Adam:2015rua,Akrami:2018vks}. However, there still exist some phenomena which are inconsistent with the cosmological principle, such as the alignment of quasar polarization vectors on large scale \cite{Hutsemekers:2005iz}, the spatial variation of the fine structure constant \cite{Webb:2010hc,King:2012id} and MOND acceleration scale \cite{Zhou:2017lwy,Chang:2018vxs,Chang:2018lab}, the anisotropic accelerating expansion of the Universe \cite{Bonvin:2006en,Antoniou:2010gw,Koivisto:2008ig,Chang:2014wpa,Chang:2014nca,Lin:2016jqp}, the alignment of CMB quadrupole and octopole \cite{Tegmark:2003ve,Bielewicz:2004en,Copi:2013jna}, the hemispherical power asymmetry in CMB \cite{Eriksen:2003db,Hansen:2004vq,Ade:2013nlj,Ade:2015hxq}, and parity asymmetry in CMB \cite{Ade:2013nlj,Ade:2015hxq,Kim:2010gd,Kim:2010gf,Zhao:2013jya,Gruppuso:2010nd}. In particular, hemispherical power asymmetry, initially observed in WMAP \cite{Bennett:2012zja,Hinshaw:2012aka}, has come to be one of the outstanding anomalies that indicated violation of statistical isotropy on large angular scales of CMB sky. This anisotropic signal persisted in Planck \cite{Ade:2013nlj,Ade:2015hxq}. This hemispherical power asymmetry has been modeled as a dipole modulation of otherwise statistically isotropic CMB sky, $\Delta \tilde{T}(\hat{\bm n})=\Delta T(\hat{\bm n})(1+A\hat{\bm \lambda} \cdot \hat{\bm n})$, where $\Delta \tilde{T}(\hat{\bm n})$ is the modulated/observed CMB field in the direction $\hat{\bm n}$, and $\Delta T(\hat{\bm n})$ is the isotropic CMB field in the same direction. $A$ and $\hat{\bm \lambda}$ are the amplitude and direction of modulation. The recent released data of Planck collaboration show deviations from isotropy with a level of significance ($\sim3\sigma$) \cite{Ade:2013nlj}.
These phenomena may imply the existence of cosmic anisotropy.

As standard candles \cite{Riess:1998cb,Perlmutter:1998np}, the supernovae of type Ia (SNe Ia) have been used in a number of works to examine the cosmological principle \cite{Antoniou:2010gw,Chang:2014wpa,Chang:2014nca,Lin:2016jqp,Schwarz:2007wf,Gupta2008,Blomqvist:2010ky,Colin:2010ds,Cai:2011xs,Mariano:2012wx,Cai:2013lja,Kalus:2012zu,Zhao:2013yaa,Wang:2014vqa,Yang:2013gea,Chang:2014jza,Heneka:2013hka,Bengaly:2015dza,Li:2015uda,Javanmardi:2015sfa,Lin:2015rza,Salehi:2015ira,Salehi:2016sta,Li:2017rqj,Ghodsi:2016dwp,Wang:2017ezt,Andrade:2017iam,Chang:2017bbi,Deng:2018yhb}. In these studies, the most commonly used datasets are given by the Union2 sample \cite{Amanullah:2010vv}, Union2.1 sample \cite{Suzuki:2011hu} and ``Joint Light-curve Analysis'' (JLA) compilation \cite{Betoule:2014frx}. Certain preferred directions were found in the Union2 sample by the hemisphere comparison method \cite{Antoniou:2010gw,Chang:2014wpa,Chang:2014nca,Lin:2016jqp,Schwarz:2007wf,Cai:2011xs,Kalus:2012zu,Yang:2013gea,Bengaly:2015dza}. The dark energy dipole was found at 2$\sigma$ level in the Union2 sample \cite{Mariano:2012wx}. Zhao et al. \cite{Zhao:2013yaa} found a dipole of deceleration parameter at more than 2$\sigma$ level by dividing the Union2 sample into 12 subsets. A dataset composed of the SNe Ia with $z<0.5$ in the Union2 was showed to deviate from the $\rm{\Lambda}$CDM model at $2\sigma\sim3\sigma$ level \cite{Colin:2010ds}. Different from the Union2 sample, the JLA sample doesn't give any convincing signal of deviation from the isotropic universe. Wang et al. \cite{Wang:2017ezt} used the JLA sample to constrain the anisotropic universe model with Bianchi-I metric and found the model was consistent with the isotropic universe. Constraining the anisotropic amplitude and direction in three different cosmological models of dark energy with the JLA sample gave a zero result \cite{Lin:2015rza}. Sang et al. \cite{Chang:2017bbi} performed a tomographic analysis on the JLA sample taking account of redshift dependence of SNe Ia color-luminosity parameter $\beta$ in the dipole-modulated $\rm{\Lambda}$CDM model, but they did not find any significant deviation from the isotropic universe. 

Recently, the Pantheon supernovae sample \cite{Scolnic:2017caz} has been released, which consists 1048 spectroscopically confirmed SNe Ia covering the redshift range $0.01<z<2.26$. Compared to the Union2 and JLA sample, the number of SNe Ia in the Pantheon sample is enlarged. The distribution of SNe Ia in Pantheon are inhomogeneous and half of them are located at south-east of the galactic coordinate system. The systematic uncertainties have been reduced by the cross calibration between subsamples in Pantheon. Therefore, the Pantheon sample could bring much stronger constraint on the anisotropy of the Universe. Previously, the Pantheon sample had been used to test the cosmological principle. Sun et al. \cite{Sun:2018cha} used a redshift tomography method to investigate the cosmic anisotropy in the Pantheon sample, and found that the isotropic cosmological model is an excellent approximation. No evidence of the cosmic anisotropy was found in the Pantheon sample by using the hemisphere comparison method, the dipole fitting method and the HEALPix \cite{Deng:2018jrp}. Zhao et al. \cite{Zhao:2019azy} investigate the cosmic anisotropy by five combinations among Pantheon, and found that the Low-$z$ and SNLS subsamples have decisive impact on the hemisphere anisotropy while the SDSS subsample has decisive impact on the dipole anisotropy. All these tests are based on the $\rm{\Lambda}$CDM cosmological model. We want to see whether the cosmic anisotropy appear in other cosmological model? In this paper, the Pantheon sample is used to constrain the possible dipole anisotropy of four cosmological models, which include the Finslerian cosmological model, the dipole-modulated $\rm{\Lambda}$CDM, $w$CDM and Chevallier--Polarski--Linder (CPL) model. We will use the MCMC method to explore the whole parameter space and find out the best fitting parameter.

The rest of this paper is arranged as follows. In section \ref{sec:data and methodology}, we briefly introduce the Pantheon sample and  four anisotropic cosmological models. In section \ref{sec:results}, we show the constraints on the four anisotropic cosmological models. Finally, conclusions are given in section \ref{sec:conclusion}.

\section{Methodology}\label{sec:data and methodology}
In the spatially flat spacetime, the distance modulus is defined as
\begin{equation}
\label{lumdis}
\mu_{t h}=5 \log _{10} \frac{d_{L}}{\mathrm{Mpc}}+25,
\end{equation}
where the luminosity distance is given as $d_{L}=\left(c / H_{0}\right) D_{L}$, $H_0$ is the Hubble constant, $c$ is the speed of light and $D_{L}$ takes the form,
\begin{equation}
D_{L}=\left(1+z_{cmb}\right) \int_{0}^{z_{c m b}} \frac{d z}{E(z)},
\end{equation} 
where $z_{cmb}$ denotes CMB frame redshift. The expression of $E(z)$ varies with different cosmological models. In the $\rm{\Lambda}$CDM model, $E(z)$ is given as
\begin{equation}
E^2(z)=\Omega_{m}(1+z)^3+(1-\Omega_{m}),
\end{equation}
where $\Omega_m$ is the matter density at the present epoch. In the $w$CDM model, $E(z)$ is given as
\begin{equation}
E^2(z)=\Omega_{m}(1+z)^{3}+\left(1-\Omega_{m}\right)(1+z)^{3(1+w_0)},
\end{equation}
where $w_0=p/\rho$ is the equation of state of dark energy. If $w_0=-1$, the $w$CDM model reduces to $\rm{\Lambda}$CDM model. In the CPL parameterization \cite{Chevallier:2000qy,Linder:2002et}, the equation of state of dark energy is redshift-dependent, $E(z)$ takes the form
\begin{eqnarray}
\nonumber
E^2(z)&=&\Omega_{m}(1+z)^{3}+\left(1-\Omega_{m}\right)(1+z)^{3\left(1+w_{0}+w_{1}\right)}\\
&&\times \exp \left(-3w_{1}\frac{z}{1+z}\right).
\end{eqnarray}
The CPL model reduces to $w$CDM model when $w_1 =0$.

\begin{figure*}
	\begin{center}
		\includegraphics[width=0.75\textwidth]{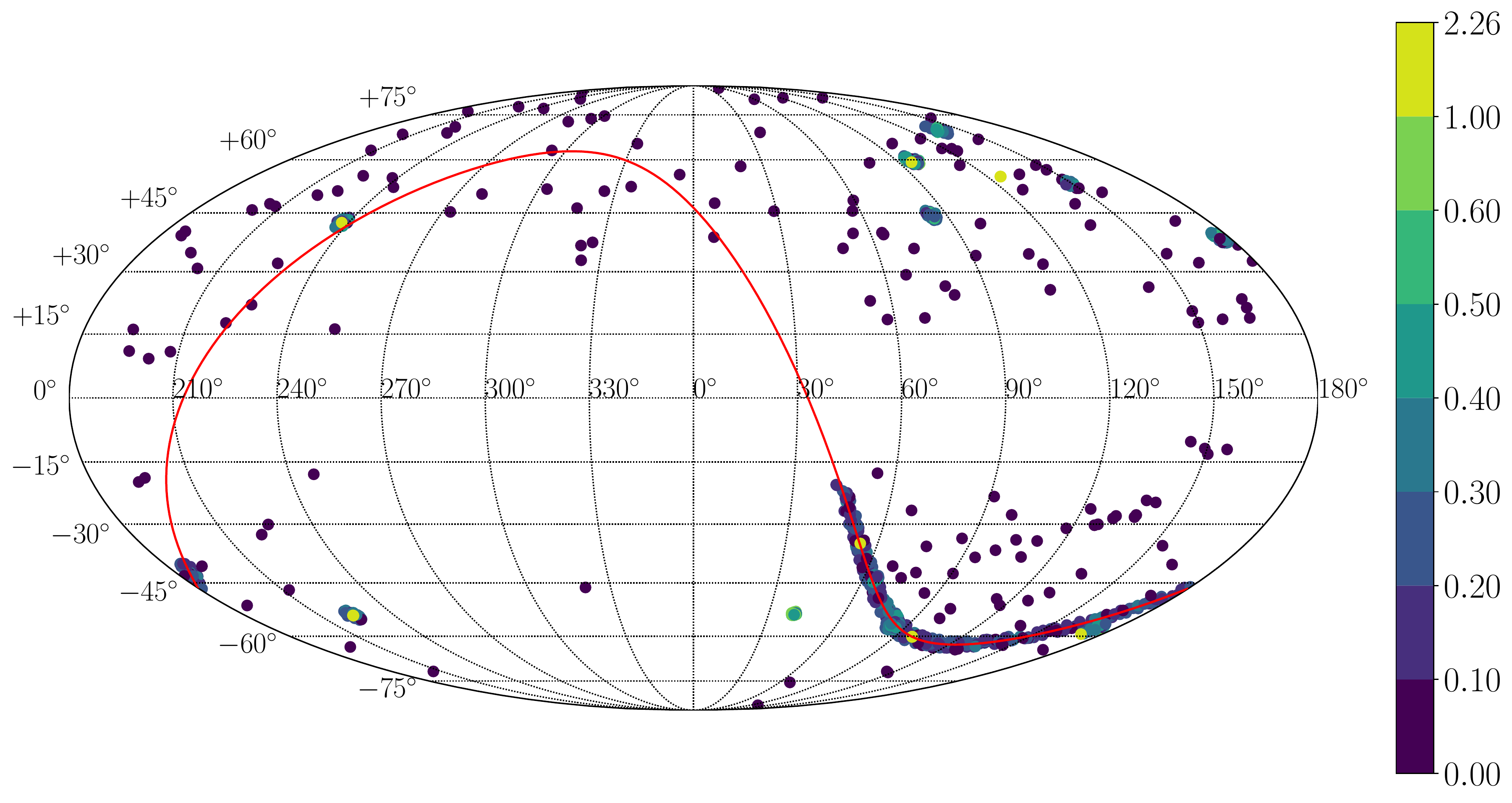}
		\caption{\label{SNIa_d}  The distribution of 1048 SNe Ia in the galactic coordinate system. The pseudo-colors indicate the redshift of these SNe Ia. The red solid curve represents the celestial equator.}
	\end{center}
\end{figure*}

Different with the $\rm{\Lambda}$CDM model, $w$CDM model and CPL model, there exist a preferred direction in the Finsler spacetime which breaks the isotropy of the Universe \cite{Chang:2014wpa,Chang:2014jza,Li:2015uda,Chang:2013xwa}. The cosmic anisotropy could be originated from the anisotropic background spacetime. Finsler spacetime admits less symmetry than the Riemann one does, which is a possible candidate for investigating the cosmological preferred direction and the dipole structure \cite{Book by Bao,Deng,Pfeifer1,Finsler PF}. We previously proposed a Finsler spacetime scenario of the anisotropic universe, which gives a unified description for dipoles of the fine-structure constant and SNe Hubble diagram \cite{Li:2015uda}. It is interesting to test this Finsler spacetime scenario by the new released SNe data. In that specific Finsler spacetime, i.e., the Randers spacetime, the scale factor have form $a=(1+A_D\cos\theta)/(1+z)$, where $A_D$ is a parameter of Randers metric, which can be regarded as the dipole amplitude. When $A_D=0$, the Randers metric would reduce to the FLRW metric, and the Finslerian cosmological model reduces to $\rm{\Lambda}$CDM model. More details about the Finsler spacetime scenario could be found therein \cite{Li:2015uda}. Correspondingly, then $E(z)$ has the form
\begin{equation}
E^2(z)=\Omega_{m}(1+z)^3(1-3A_D \cos \theta)+(1-\Omega_{m}),
\end{equation}
where $\theta$ is the angle between the preferred direction in the Finsler spacetime and the position of the SNe Ia. In the galactic coordinate system, the parameterization of Finslerian preferred direction is the same with the dipole direction $\bm{\hat{n}}$ discussed later in Eq. (\ref{nn}).

In this paper, we use the recently released ``Pantheon" sample to constrain the possible dipole anisotropy in the four cosmological models mentioned above. The Pantheon sample consists 1048 SNe Ia in the redshift range of 0.01 to 2.26. It is a collection of SNe Ia discovered by the Pan-STARRS1 (PS1) Medium Deep Survey and SNe Ia from Low-$z$, SDSS, SNLS and HST surveys. Compared to Union2 and JLA sample, the number of SNe Ia in the Pantheon sample is enlarged and the systematic uncertainties have been reduced by the cross calibration between subsamples. Fig. \ref{SNIa_d} shows the distribution of 1048 SNe Ia in the galactic coordinate system. As we can see, the distribution of these SNe Ia is inhomogeneous and half of them are located at south-east of the galactic coordinate system. Especially, there are 335 SNe Ia clustering in a narrow strip which corresponds to the equator of the equatorial coordinate system, that could bring significant impact on the cosmic anisotropy. 

In the Pantheon sample, the observed distance module is determined by a modified version of the Tripp formula \cite{Tripp:1998},
\begin{equation}
\mu_{o b s}=m_{B}-M+\alpha x_{1}-\beta c+\Delta_{M}+\Delta_{B},
\end{equation}
where $\mu_{obs}$ denotes the observed distance modulus, $m_B$ and $M$ are the apparent magnitude and absolute magnitude of SNe Ia in B-band, respectively. $x_1$ is the stretch parameter and $c$ is the color parameter. $\alpha$ represents the coefficient of the relation between luminosity and stretch. $\beta$ represents the coefficient of the relation between luminosity and color. $\Delta_M$ and $\Delta_B$ are distance corrections depending on the host galaxy mass of the SNe Ia and the predicted biases from simulations, respectively. Since $M$ is degenerated with $H_0$, Scolnic et al. \cite{Scolnic:2017caz} gave a corrected apparent magnitudes, i.e., $m_{o b s}=\mu_{o b s}+M$. The theoretical apparent magnitude is given as
\begin{equation}
\label{mm}
m_{t h}=\mu_{t h}+M=5 \log _{10} D_{L}+\mathcal{M},
\end{equation}
where $\mathcal{M}$ is an nuisance parameter, which depends on the Hubble constant $H_0$ and the absolute  magnitude $M$. 

The dipole fitting method, which proposed by Mariano $\&$ Perivolaropoulos \cite{Mariano:2012wx}, is widely used to investigate the anisotropy of the Universe. In this paper, we consider a dipole modulation to the theoretical apparent magnitude in the isotropic $\rm{\Lambda}$CDM model, $w$CDM model and CPL model, namely
\begin{equation}
\widetilde{m}_{th}=m_{th}[1+A_D(\bm{\hat{n}} \cdot \bm{\hat{p}})],
\end{equation} 
where $A_D$ indicates the dipole amplitude, $\bm{\hat{n}}$ is the direction of dipole and $\bm{\hat{p}}$ is the unit vector pointing to the SNe Ia. ${m}_{th}$ indicates the theoretical apparent magnitude in the isotropic $\rm{\Lambda}$CDM model, $w$CDM model and CPL model given by Eq. (\ref{mm}). In the galactic coordinate system, the direction of dipole $\bm{\hat{n}}$ can be parameterized as
\begin{equation}\label{nn}
\bm{\hat{n}}=\cos(b)\cos(l)\bm{\hat{i}}+\cos(b)\sin(l)\bm{\hat{j}}+\sin(b)\bm{\hat{k}},
\end{equation}
where $l$ is the galactic longitude and $b$ is the galactic latitude. $\bm{\hat{i}}$, $\bm{\hat{j}}$ and $\bm{\hat{k}}$ are unit vectors along the axis in the Cartesian coordinates system. Similarly, the position of the $i$th SNe Ia can be parameterized as
\begin{equation}
\bm{\hat{p}_i}=\cos(b_i)\cos(l_i)\bm{\hat{i}}+\cos(b_i)\sin(l_i)\bm{\hat{j}}+\sin(b_i)\bm{\hat{k}}.
\end{equation}
Then we can compare the corrected apparent magnitudes $m_{o b s}$ and the dipole-modulated theoretical apparent magnitude $\widetilde{m}_{th}$ to constrain the parameters in three dipole-modulated cosmological models. For the Finslerian cosmological model, the spacetime is intrinsic anisotropic and the dipole modulation is unnecessary, and the parameters are constrained by the comparison of $m_{o b s}$ and $m_{th}$.

To explore the whole cosmological parameter space, we employ the $\chi^2$ statistic,
\begin{equation}
\chi^2=\Delta  \bm{\mu}^T \cdot \bm{C}^{-1} \cdot \Delta \bm{\mu}=\Delta \bm{m}^T \cdot \bm{C}^{-1} \cdot \Delta \bm{m},	
\end{equation}
where $\Delta \bm{\mu}=\bm{\mu}_{obs}-\bm{\mu}_{th}$ or $\Delta \bm{m}=\bm{m}_{obs}-\bm{m}_{th}$. The total covariance matrix takes the form
\begin{equation}
\bm{C}=\bm{D}_{stat}+\bm{C}_{sys},
\end{equation}
where the diagonal matrix $\bm{D}_{stat}$ and the covariance matrix $\bm{C}_{sys}$ denotes the statistical uncertainties and the systematic uncertainties, respectively.

\emph{Note added}: The Pantheon supernovae data has been updated in GitHub repository\footnote{\url{https://github.com/dscolnic/Pantheon}}. $z_{cmb}$ has been corrected by the peculiar velocity correction for $z<0.08$ in updated file {\tt lcparam$_-$full$_-$long$_-$zhel.txt}. The corrected apparent magnitudes  $m_{o b s}$ and its statistical uncertainties are also given in that updated file. The systematic uncertainties are given in the file {\tt sys$_-$full$_-$long.txt}. In addition, the position of each SNe Ia could be found in the folder {\tt data$_-$fitres}.

\section{Results}\label{sec:results}

\begin{figure*}
	\begin{center}
		\includegraphics[width=0.9\textwidth]{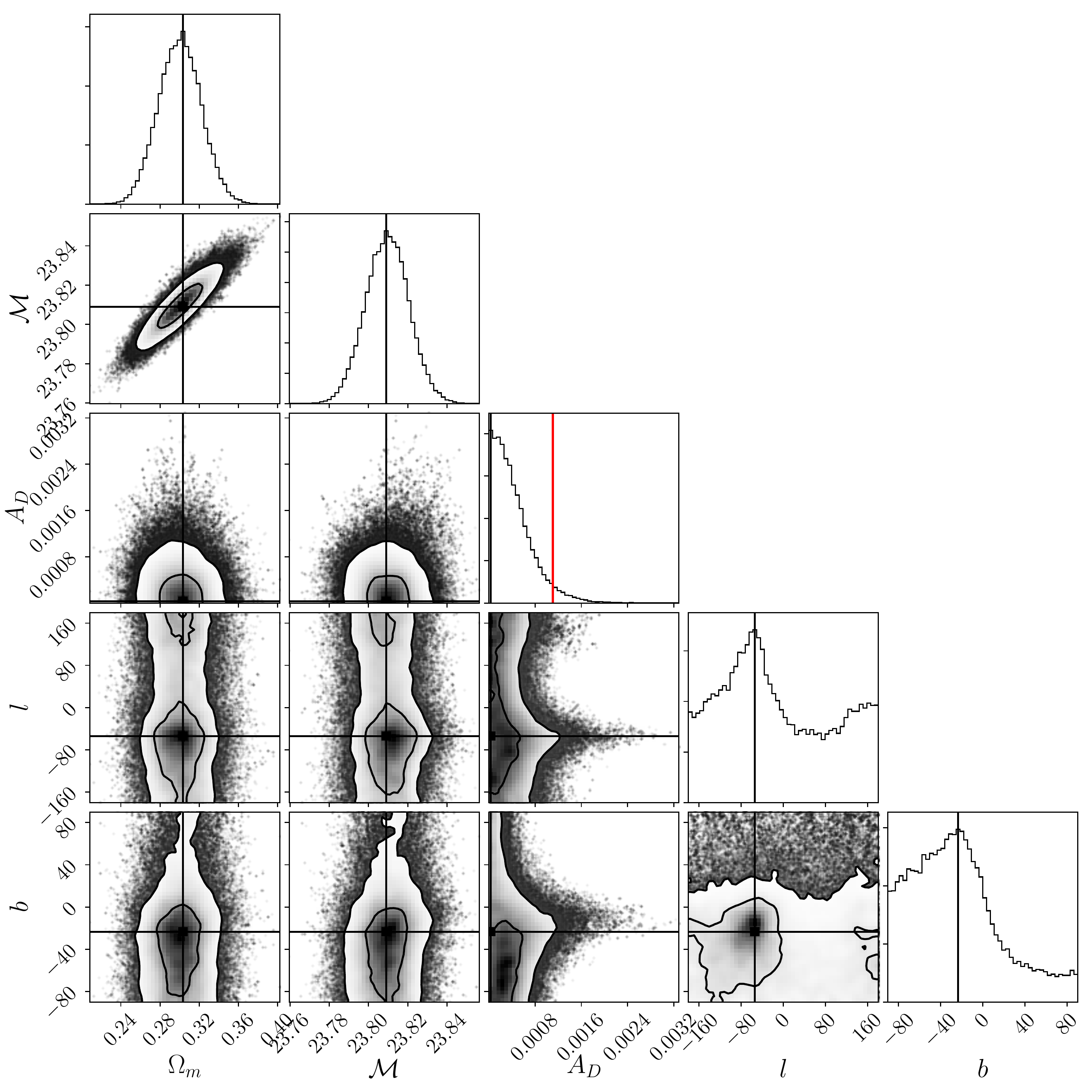}
		\caption{The 1-dimensional and 2-dimensional marginalized posterior distributions for the parameter space in the dipole-modulated $\rm{\Lambda}$CDM model. The horizontal and vertical solid black lines mark the maximum of 1-dimensional marginalized posteriors. The red line indicates the $95\%$ CL upper limit of dipole amplitude $A_D$.  $l$ and $b$ is in the unit of degree.}
		\label{LCDM}
	\end{center}
\end{figure*}

\begin{figure*}
	\begin{center}
		\includegraphics[width=0.9\textwidth]{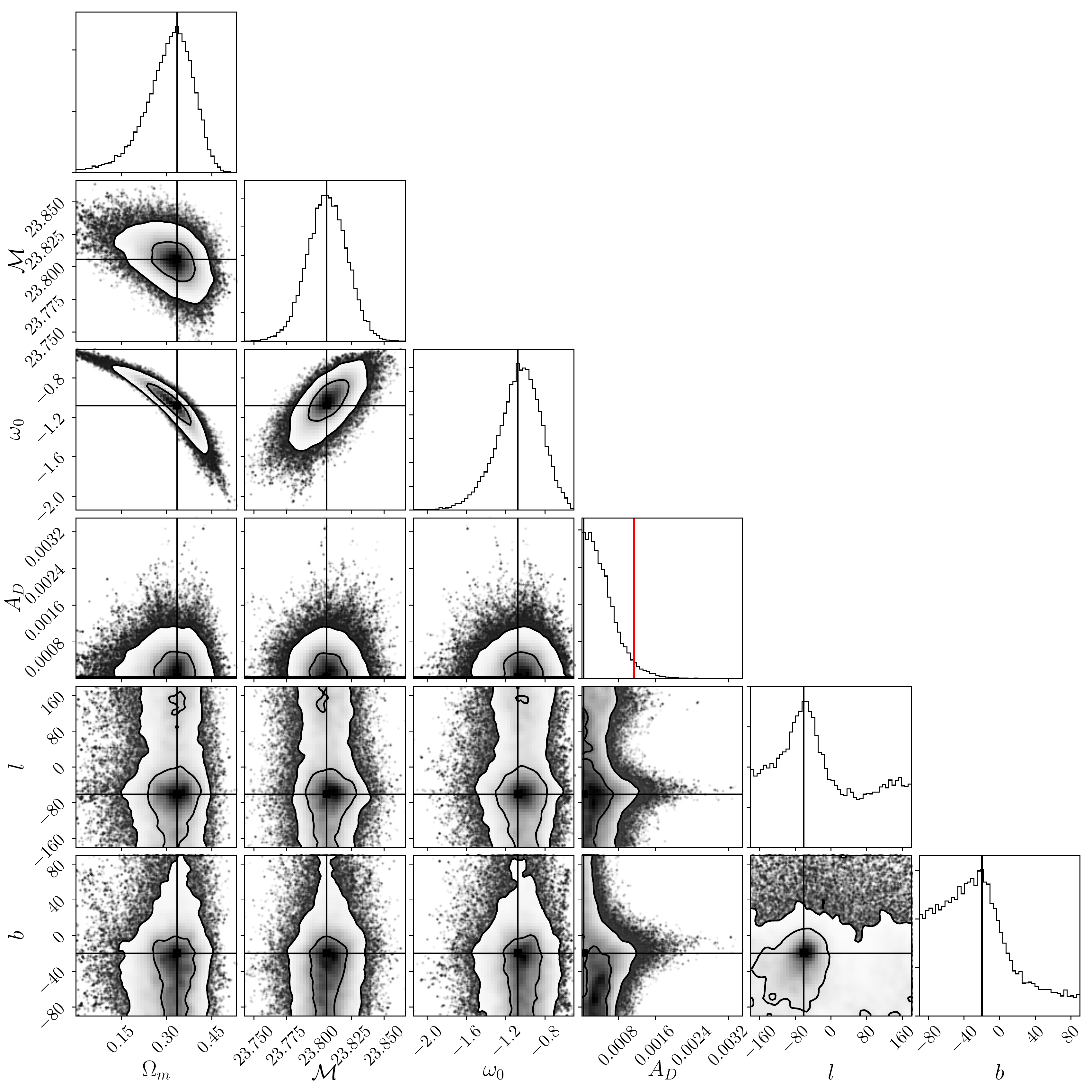}
		\caption{The 1-dimensional and 2-dimensional marginalized posterior distributions for the parameter space in the dipole-modulated $w$CDM model. The horizontal and vertical solid black lines mark the maximum of 1-dimensional marginalized posteriors. The red line indicates the $95\%$ CL upper limit of dipole amplitude $A_D$.  $l$ and $b$ is in the unit of degree.}
		\label{wCDM}
	\end{center}
\end{figure*}

\begin{figure*}
	\begin{center}
		\includegraphics[width=0.9\textwidth]{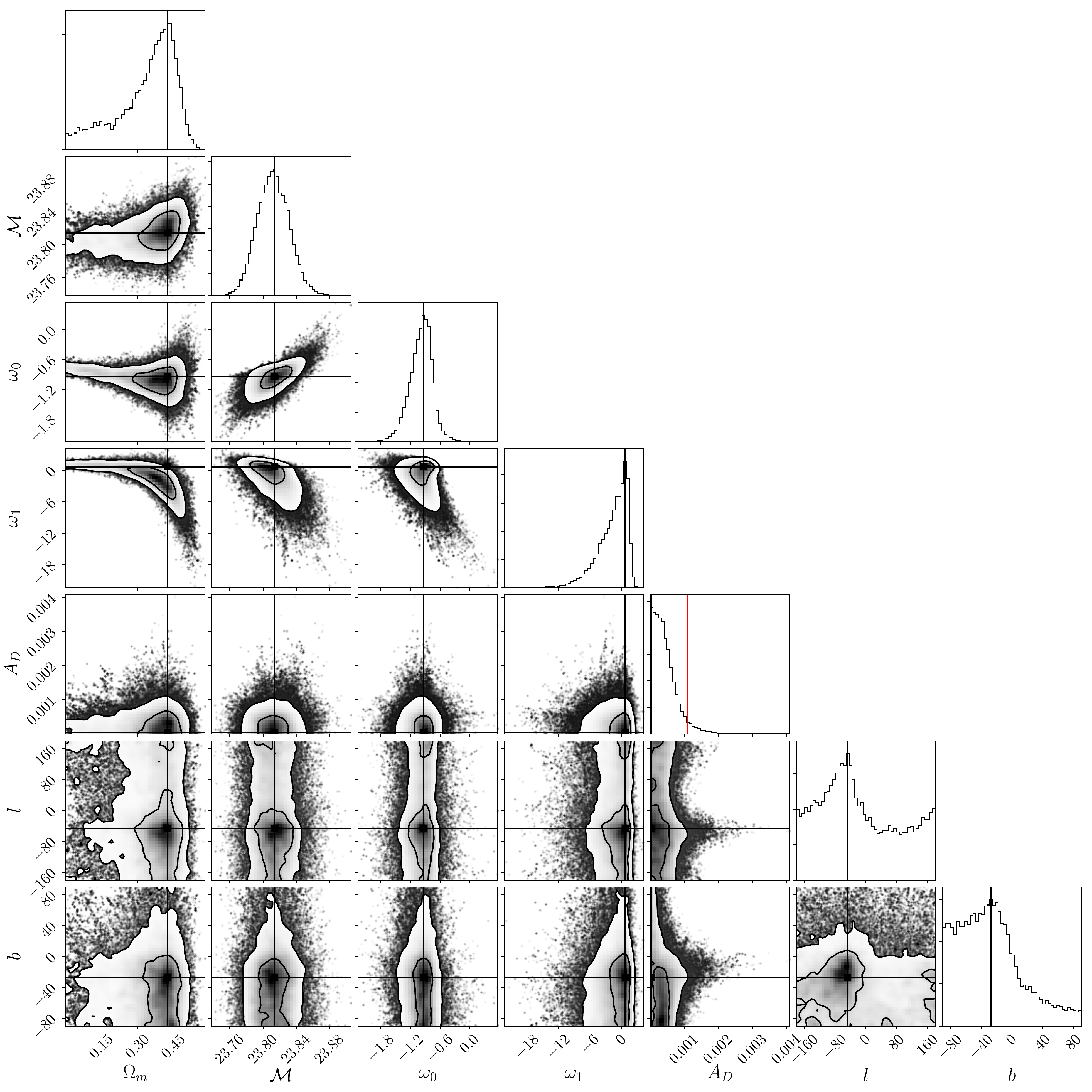}
		\caption{The 1-dimensional and 2-dimensional marginalized posterior distributions for the parameter space in the dipole-modulated CPL model. The horizontal and vertical solid black lines mark the maximum of 1-dimensional marginalized posteriors. The red line indicates the $95\%$ CL upper limit of dipole amplitude $A_D$.  $l$ and $b$ is in the unit of degree.}
		\label{CPL}
	\end{center}
\end{figure*}

\begin{figure*}
	\begin{center}
		\includegraphics[width=0.9\textwidth]{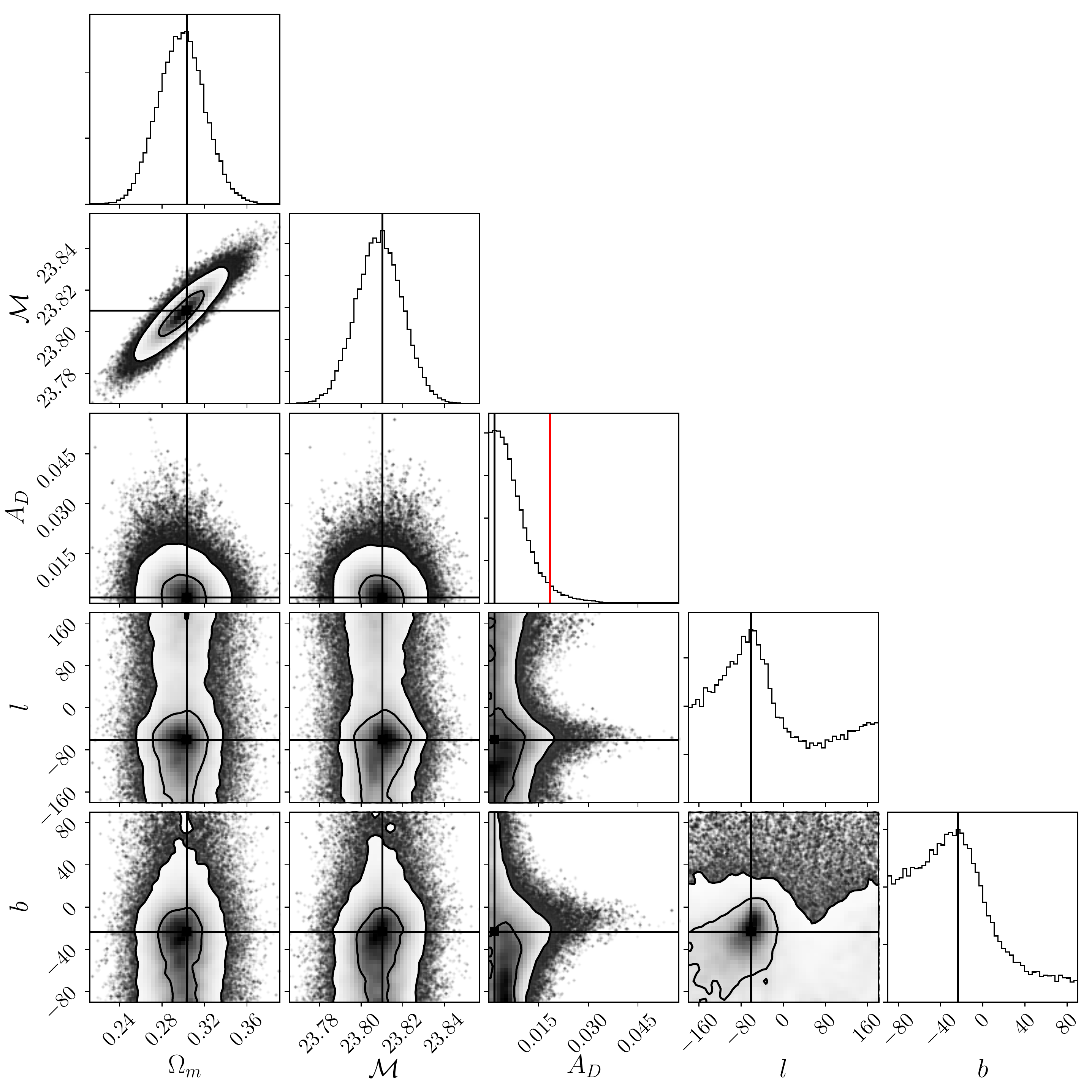}
		\caption{The 1-dimensional and 2-dimensional marginalized posterior distributions for the parameter space in the Finslerian cosmological model. The horizontal and vertical solid black lines mark the maximum of 1-dimensional marginalized posteriors. The red line indicates the $95\%$ CL upper limit of dipole magnitude $A_D$.  $l$ and $b$ is in the unit of degree.}
		\label{Finsler}
	\end{center}
\end{figure*}

\begin{table*}
	\begin{center}
		\caption{The best fitting values for three dipole-modulated cosmological models as well as Finslerian cosmological model. We show the maximum and its $68\%$ CL constraints on the model parameters $\Omega_m$, $\mathcal{M}$, $w_0$, $w_1$, $l$, $b$, and the $95\%$ CL upper limits of dipole amplitude $A_D$.}
		\setlength{\tabcolsep}{3mm}{
			\begin{tabular}{lccccccc}
				\hline
				\multicolumn{1}{l}{Model}               &$\Omega_m$                   &$\mathcal{M}$                  &$w_0$                         &$w_1$                        &$A_D[10^{-3}]$ &$l[^{\circ}]$                  &$b[^{\circ}]$                 \\
				\hline
				\multicolumn{1}{l}{$\rm{\Lambda}$CDM}   &$ 0.303_{ -0.025}^{ +0.019}$ &$ 23.809_{ -0.010}^{ +0.011}$  &$-$                           &$-$                          &$ <1.11$       &$ 306.00_{ -125.98}^{ +91.94}$ &$ -23.41_{ -54.71}^{ +22.97}$ \\
				\multicolumn{1}{l}{$w$CDM}              &$ 0.335_{ -0.082}^{ +0.066}$ &$ 23.806_{ -0.014}^{ +0.016}$  &$ -1.079_{ -0.162}^{ +0.271}$ &$-$                          &$ <1.14$       &$ 298.81_{ -118.71}^{ +84.18}$ &$ -19.80_{ -63.25}^{ +14.07}$ \\
				\multicolumn{1}{l}{CPL}                 &$ 0.423_{ -0.141}^{ +0.060}$ &$ 23.814_{ -0.020}^{ +0.018}$  &$ -0.937_{ -0.215}^{ +0.218}$ &$ 0.711_{ -3.633}^{ +0.959}$ &$ <1.09$       &$ 313.20_{ -133.15}^{ +75.30}$ &$ -27.00_{ -57.24}^{ +18.72}$ \\
				\multicolumn{1}{l}{Finslerian}          &$ 0.303_{ -0.027}^{ +0.017}$ &$ 23.810_{ -0.013}^{ +0.009}$  &$-$                           &$-$                          &$<18.50$       &$ 298.80_{ -118.69}^{ +75.31}$ &$ -23.41_{ -57.41}^{ +19.26}$ \\
				\hline
		\end{tabular}}
		\label{table:DF1}
	\end{center}
\end{table*}

\begin{figure*}
	\begin{center}
		\includegraphics[width=0.75\textwidth]{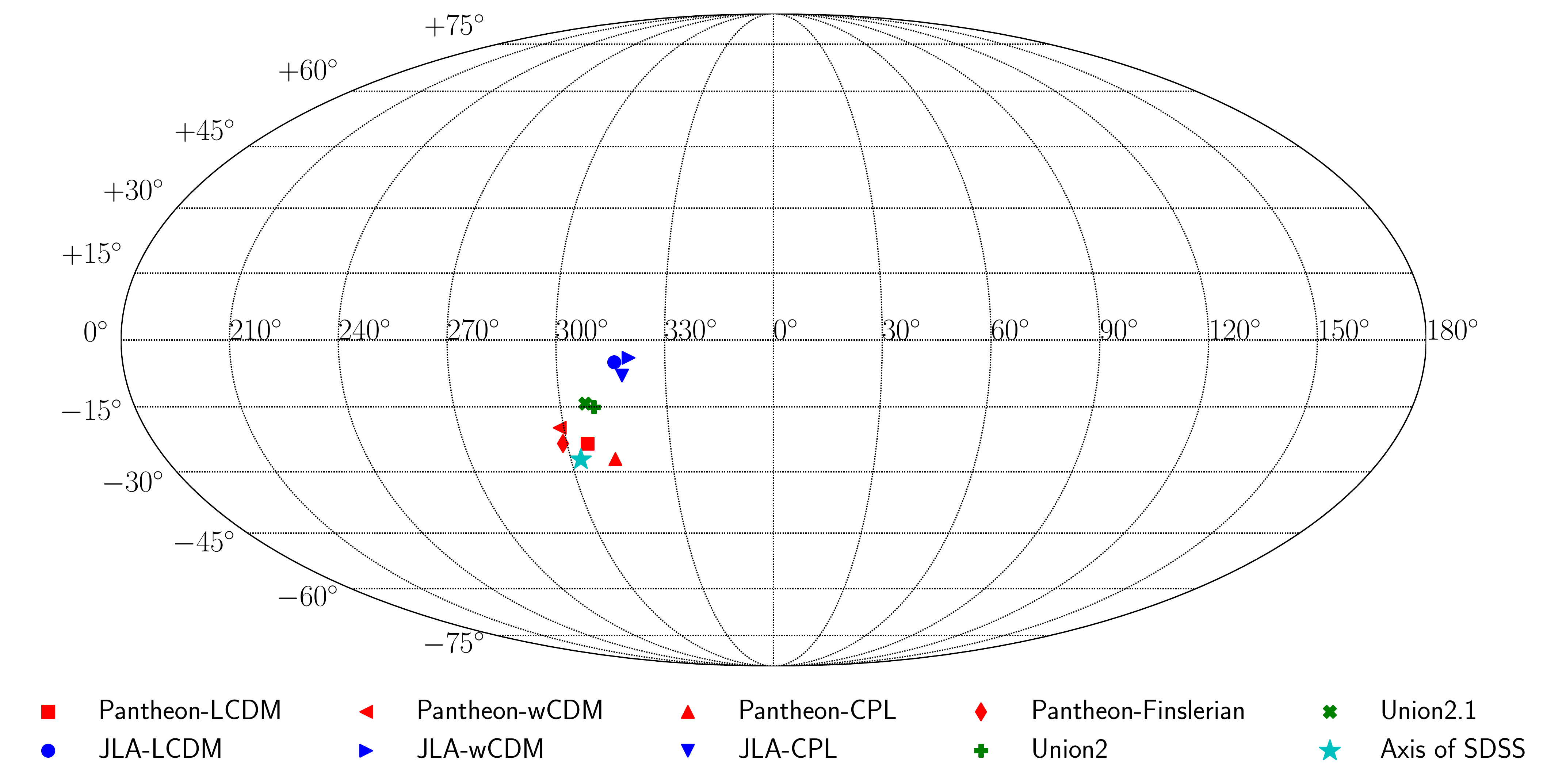}
		\caption{The dipole directions in four cosmological models by using the Pantheon sample, and the dipole directions derived from the Union2 sample \cite{Mariano:2012wx}, Union2.1 sample \cite{Yang:2013gea} and JLA sample \cite{Lin:2015rza}. The star marks the axial direction to the plane of SDSS subsample \cite{Zhao:2019azy}. The uncertainties of dipole directions could be found in Table \ref{table:DF1}.}
		\label{DF}
	\end{center}
\end{figure*}

In this paper, we use the MCMC method to explore the whole parameter space. Specifically, we use the affine-invariant MCMC ensemble sampler provided by \emph{emcee}\footnote{\url{https://emcee.readthedocs.io/en/stable/}} \cite{ForemanMackey:2012ig} , which is widely used in astrophysics and cosmology. In the dipole-modulated $\rm{\Lambda}$CDM model, the fitting parameters consist of the matter density $\Omega_m$, the nuisance parameter $\mathcal{M}$, the dipole amplitude $A_D$ and the dipole direction ($l$, $b$). Compared with dipole-modulated $\rm{\Lambda}$CDM model, the dipole-modulated $w$CDM model has an extra parameter $w_0$ and the dipole-modulated CPL model has two extra parameter $w_0$ and $w_1$. For the Finslerian cosmological model, formally, it has the same parameter space with the dipole-modulated $\rm{\Lambda}$CDM model, but the dipole anisotropy is derived from a specific Finsler spacetime.
In the MCMC method, the posterior distributions are determined by priors and likelihood functions, and the latter is given as $\mathcal{L} \propto \exp \left(-\chi^{2} / 2\right)$.
We use flat prior on each parameter as follow: $\Omega_m\sim[0,1],~\mathcal{M}\sim[0,100],~w_0\sim[-100,100],~w_1\sim[-100,100],~A_D\sim[0,1],~l\sim[-180^\circ,180^\circ],b\sim[-90^\circ,90^\circ]$.

As mentioned above, we constrain the dipole anisotropy in four cosmological models, i.e., the dipole-modulated $\rm{\Lambda}$CDM, $w$CDM, CPL model and the Finslerian cosmological model by using the Pantheon sample. The best fitting parameters can be derived by maximizing the posterior. Our results are shown in Fig. \ref{LCDM}-\ref{Finsler} and summarized in Table \ref{table:DF1}. In Fig. \ref{LCDM}-\ref{Finsler}, we show the marginalized posterior distribution for each cosmological model. In Table \ref{table:DF1}, we show the 95\% confidence level (CL) upper limit of the dipole amplitude $A_D$, the maximum and the 68\% CL constraints\footnote{We use the highest posterior density (HPD) credible interval method to determine the 68\% CL.} on other parameters. Note that the galactic longitude has been converted to positive value. 

In the dipole-modulated $\rm{\Lambda}$CDM model, the matter density $\Omega_m$ and the nuisance parameter $\mathcal{M}$ are well constrained by Pantheon sample. The results are $\Omega_m= 0.303_{ -0.025}^{ +0.019}$ and $\mathcal{M}=23.809_{ -0.010}^{ +0.011}$, which is in agreement with the results in Scolnic et al. \cite{Scolnic:2017caz}, and they did not consider the dipole anisotropy. In our case, we find the dipole anisotropy is very weak, and the dipole amplitude is constrained as $A_D < 1.11\times 10^{-3}$ at 95\% CL and the dipole direction points towards $(l,b)=({306.00^{\circ}}_{-125.98^{\circ}}^{+91.94^{\circ}},{-23.41^{\circ}}_{-54.71^{\circ}}^{+22.97^{\circ}})$.	The very large uncertainty of dipole direction also implies that the dipole anisotropy in the dipole modulated $\rm{\Lambda}$CDM model is very weak.

In the dipole-modulated $w$CDM model, the extra parameter, i.e., the equation of state of dark energy is $w_0=-1.079_{ -0.162}^{ +0.271}$, which is also consistent with the results in Scolnic et al. \cite{Scolnic:2017caz}. The matter density is constrained as $\Omega_m=0.335_{ -0.082}^{ +0.066}$ and the nuisance parameter is constrained as $\mathcal{M}=23.806_{ -0.014}^{ +0.016}$. As same as the dipole-modulated $\rm{\Lambda}$CDM model, the dipole anisotropy is very weak and the dipole amplitude is constrained to be $A_D < 1.14\times 10^{-3}$ at $95\%$ CL. The dipole direction points towards $(l,b)=({ 298.81^{\circ}}_{-118.71^{\circ}}^{ +84.18^{\circ}},{-19.80^{\circ}}_{-63.25^{\circ}}^{+14.07^{\circ}})$, which is very close to the dipole direction in the dipole-modulated $\rm{\Lambda}$CDM model.

In the dipole-modulated CPL model, there are two extra parameters $w_0$ and $w_1$ which are constrained as $w_0=-0.937_{ -0.215}^{ +0.218}$ and $w_1= 0.711_{ -3.633}^{ +0.959}$. The matter density is $\Omega_m=0.423_{ -0.141}^{ +0.060}$, which is slightly larger than that in the dipole-modulated $\rm{\Lambda}$CDM or $w$CDM model and the nuisance parameter is $\mathcal{M}=23.814_{ -0.020}^{ +0.018}$. The dipole amplitude is constrained to be $A_D < 1.09\times 10^{-3}$ at $95\%$ CL. The dipole direction points towards $(l,b)=({313.20^{\circ}}_{-133.15^{\circ}}^{+75.30^{\circ}},{-27.00^{\circ}}_{-57.24^{\circ}}^{+18.72^{\circ}})$, which is very close to the dipole directions mentioned above.

In the Finslerian cosmological model, the matter density $\Omega_m$ and the nuisance parameter $\mathcal{M}$ are constrained as $\Omega_m=0.303_{ -0.027}^{ +0.017}$ and $\mathcal{M}=23.810_{ -0.013}^{ +0.009}$ , which is almost the same with that in the dipole-modulated $\rm{\Lambda}$CDM model. The dipole anisotropy is very weak, and the dipole amplitude is constrained as $A_D < 18.50 \times 10^{-3}$ at $95\%$ CL. The dipole direction points towards $(l,b)=({298.80^{\circ}}_{-118.69^{\circ}}^{+75.31^{\circ}},$
${-23.41^{\circ}}_{-57.41^{\circ}}^{+19.26^{\circ}})$, which is very close to the dipole directions in other three dipole-modulated cosmological models.

At the end, we make some comparisons between the dipole directions mentioned above in the Pantheon sample with that derived from the Union2 sample \cite{Mariano:2012wx}, Union2.1 sample \cite{Yang:2013gea} and JLA sample \cite{Lin:2015rza}. The dipole directions in each SNe Ia sample are shown in Fig. \ref{DF}. As can be seen, the dipole directions in the four cosmological models are close to each other by using the Pantheon sample. The angular separations between these dipole directions are much smaller than its uncertainties. For JLA sample, Lin et al. \cite{Lin:2015rza} found almost the same conclusions and they considered the dipole-modulated $\rm{\Lambda}$CDM, $w$CDM and CPL models. Interestingly, all these dipole directions are close to the dipole direction in the Union2 and Union2.1 sample within $16.36^{\circ}$. This angular separation is much smaller than the uncertainty of the dipole direction. The star in Fig. \ref{DF} denotes the axial direction to the plane of SDSS subsample among Pantheon, which points towards $(l,b) = (302.93^{\circ}, -27.13^{\circ})$. Coincidentally, the dipole directions in the Pantheon sample are very close to the axial direction to the plane of SDSS subsample and the angular separation is less than $9.14^{\circ}$. Moreover, we find that the dipole directions shift more than $40.18^{\circ}$ when we exclude the SDSS subsample from Pantheon. The consistency may confirm the conclusion that the SDSS subsample plays a dominant role on the dipole anisotropy in the Pantheon sample \cite{Zhao:2019azy}. Similar conclusions was found in the Union2 sample \cite{Jimenez:2014jma}. Monte-Carlo simulations also show that the anisotropic distribution of coordinates can cause dipole directions and make dipole magnitude larger \cite{Sun:2018epo}. Therefore, we suggest that the weak dipole anisotropy in Pantheon sample may originate from the inhomogeneous distribution of the SDSS subsample among Pantheon.

\section{Conclusions}\label{sec:conclusion}
\vspace{0.1cm}
In this paper, the recently released Pantheon sample of SNe Ia was used to test the possible dipole anisotropy in the Finslerian cosmological model and other three dipole-modulated cosmological models, i.e., the dipole-modulated $\rm{\Lambda}$CDM, $w$CDM and CPL model. The MCMC method was used to explore the whole cosmological parameter space. We found that the dipole anisotropy is very weak in all cosmological models used. For the dipole-modulated $\rm{\Lambda}$CDM model, the dipole amplitude has an upper limit $1.11 \times 10^{-3}$ at $95\%$ CL and the dipole direction points towards $(l,b)=({306.00^{\circ}}_{-125.98^{\circ}}^{+91.94^{\circ}},{-23.41^{\circ}}_{-54.71^{\circ}}^{+22.97^{\circ}})$. The dipole-modulated $w$CDM and CPL models have similar dipole anisotropy. For the Finslerian cosmological model, the dipole amplitude has an upper limit $18.50 \times 10^{-3}$ at 95\% CL and the dipole direction in Finsler spacetime points towards $(l,b)=({298.80^{\circ}}_{-118.69^{\circ}}^{+75.31^{\circ}},$
\noindent
${-23.41^{\circ}}_{-57.41^{\circ}}^{+19.26^{\circ}})$. All these results show the isotropic cosmological model is an excellent approximation.
We made some comparisons and found that the dipole direction in Pantheon or JLA sample is close to the dipole direction in Union2 and Union2.1 sample. Coincidentally, these dipole directions are close to the axial direction to the plane of SDSS subsample among Pantheon. Therefore, we suggested that the weak dipole anisotropy in the Pantheon sample may originate from the inhomogeneous distribution of the SDSS subsample. More homogeneous distribution of SNe Ia is necessary to constrain the cosmic anisotropy.

\acknowledgments{We thank Zhi-Chao Zhao for useful discussions. We greatly appreciate D. M. Scolnic for private communication.} %We thank the anonymous referee for valuable suggestions and comments.

\end{multicols}
\vspace{10mm}

\begin{multicols}{2}

\end{multicols}

\clearpage

\end{CJK*}

\end{document}